\documentclass[%
 aps,
 pra,
 amsmath,amssymb,
reprint,%
superscriptaddress]
{revtex4-1}

\usepackage{graphicx}
\usepackage{dcolumn}
\usepackage{bm}
\usepackage[mathlines]{lineno}
\usepackage{amsmath}
\usepackage{url}

\usepackage{booktabs}
\usepackage{rotating}
\usepackage{multirow}
\usepackage{subfigure}
\usepackage[normalem]{ulem}
\usepackage{amsmath}
\usepackage[hyperref,usenames,dvipsnames,svgnames,table]{xcolor}
\usepackage{natbib,hyperref}
\hypersetup{
     colorlinks   = true,
     citecolor    = blue,
     linkcolor    = blue,
     urlcolor     = blue
}

\usepackage{booktabs}
\usepackage{rotating}
\usepackage{multirow}
\usepackage{subfigure}

\usepackage{color}
\usepackage[usenames,dvipsnames,svgnames,table]{xcolor}

\begin{document}
\frenchspacing

\title[Sample title]{Atomistic behavior of metal surfaces under high electric fields}

\author{A. Kyritsakis}%
\email{andreas.kyritsakis@helsinki.fi; akyritsos1@gmail.com}
\affiliation{Helsinki Institute of Physics and Department of Physics, University of Helsinki, PO Box 43 (Pietari Kalmin katu 2), 00014 Helsinki, Finland}

\author{E. Baibuz}
\affiliation{Helsinki Institute of Physics and Department of Physics, University of Helsinki, PO Box 43 (Pietari Kalmin katu 2), 00014 Helsinki, Finland}

\author{V. Jansson}%
\affiliation{Helsinki Institute of Physics and Department of Physics, University of Helsinki, PO Box 43 (Pietari Kalmin katu 2), 00014 Helsinki, Finland}

\author{F. Djurabekova}
\affiliation{Helsinki Institute of Physics and Department of Physics, University of Helsinki, PO Box 43 (Pietari Kalmin katu 2), 00014 Helsinki, Finland}

\date{\today}

\begin{abstract}
Combining classical electrodynamics and density functional theory (DFT) calculations, we develop a general and rigorous theoretical framework that describes the energetics of metal surfaces under high electric fields.
We show that the behavior of a surface atom in the presence of an electric field can be described by the polarization characteristics of the permanent and field-induced charges in its vicinity.
We use DFT calculations for the case of a W adatom on a W\{110\} surface to confirm the predictions of our theory and quantify its system-specific parameters.  
Our quantitative predictions for the diffusion of W-on-W\{110\} under field are in good agreement with experimental measurements.
This work is a crucial step towards developing atomistic computational models of such systems for long-term simulations.
\end{abstract}

\maketitle

\section{Introduction}
The interaction of metal surfaces with an applied electric field is well described in the continuum-limit by classical electrodynamics \cite{jackson1975classical}.
However, how exactly this knowledge is translated to the sub-nanometer scale in order to predict, for instance, the behavior of surface single point defects under an electric field is not yet clear.
Knowing the exact mechanisms driving the evolution of a metal surface under electric field is critical for developing various modern nanotechnologies \cite{fujita2007mechanism, yanagisawa2009optical, yanagisawa2016laser, mayer1999electric}.
Furthermore, various existing and projected devices, such as contactless atomic manipulators \cite{stroscio1991atomic, whitman1991manipulation}, electron and ion sources \cite{gomer1992field,jensen2007advances,brown2004physics,harp1990atomic,lai2017xenon}, atom probe tomography \cite{miller2014atom,miller2012atom,Perea2009direct,Kelly2007invited}, field ion and field emission microscopy \cite{muller1956field,Muller1965field,gomer1992field}, or even large-scale particle accelerators etc. \cite{Descoeudres2009,antoine2012electromigration,clic2016,engelberg2018stochastic}
would significantly benefit from the existence of such a theory.

There are many indications in the literature that metal surfaces behave differently under the influence of a high electric field \cite{Dyke1953I,Dyke1953Arc,tsong1975direct,mayer1999electric,fujita2007mechanism,parviainen2014molecular,vigonski2015molecular,Veske2016,kyritsakis2018thermal,vurpillot2018simulation}.
For example, the surface diffusion of adatoms has been found both computationally \cite{feibelman2001surface, sanchez2004field, muller2006migration} and experimentally \cite{utsugi1962field, kellogg1993electric,antczak2010surface} to vary depending on the magnitude of the applied field and even become biased when a non-uniform field is present \cite{tsong1972measurements, tsong1975direct}.
In spite of the aforementioned significance and various experimental and theoretical studies since the 1950s, the theoretical understanding of such surface-field effects on the atomistic level remains insufficient.

The behavior of adatoms on metal surfaces in the presence of an electric field has attracted interest of both theoretical and experimental studies since the 1950s \cite{becker1951use,drechsler1957ristallstufen,drechsler1960point,tsong1971measurement,tsong1972measurements,tsong1975direct,kellogg1978direct,kellogg1984measurement,feibelman2001surface,sanchez2004field}.
Tsong and Kellogg (TK) \cite{tsong1975direct} proposed a theoretical model describing this behavior in terms of the polarization characteristics of individual adatoms, which were treated as isolated neutral point dipoles.
However, this description is not compatible with a quantum mechanical picture of the metal surface, as it neglects the charge redistribution induced by the adatom in its vicinity. In addition, the adatom will not be neutral, but rather significantly charged.
Moreover, the notion of "atomic dipole moment" is fundamental to TK's model, yet it is not given a precise definition with respect to the charge distribution of the adatom-surface system.

In this letter, we present an ab initio theory, that rigorously describes the atomistic behavior of a metal surface under high electric field, in terms of the well-defined polarization characteristics of the entire surface-adatom system.
Our theory establishes a general approach for calculating the electric field effects on the activation energy of any atomic transition on a metal surface under both uniform and non-uniform electric fields, utilizing modern density functional theory (DFT) calculations.
We validate our approach by calculating the activation energies of migration and thermal evaporation for the particular case of a W adatom on a W\{110\} surface and subsequently successfully comparing our theoretical predictions to both direct DFT calculations and experimental data available in the literature.

In section \ref{sec:theory}, we develop the theoretical concepts that describe the activation energy of various atomic transitions in terms of the system polarization characteristics.
Then in section \ref{sec:method} we describe the methodology for our DFT calculations for the W\{110\} system, the results of which are presented in section \ref{sec:results}, validating our theory versus direct DFT calculations.
Finally, in section \ref{sec:discussion}, we compare our results with experiment and discuss the limits of our theory, before concluding in section

\section{Theory} \label{sec:theory}

\subsection{Dipole moment and energy} \label{sec:dipole}

\begin{figure}[!htbp]
  \centering
    \includegraphics[width=.99\linewidth]{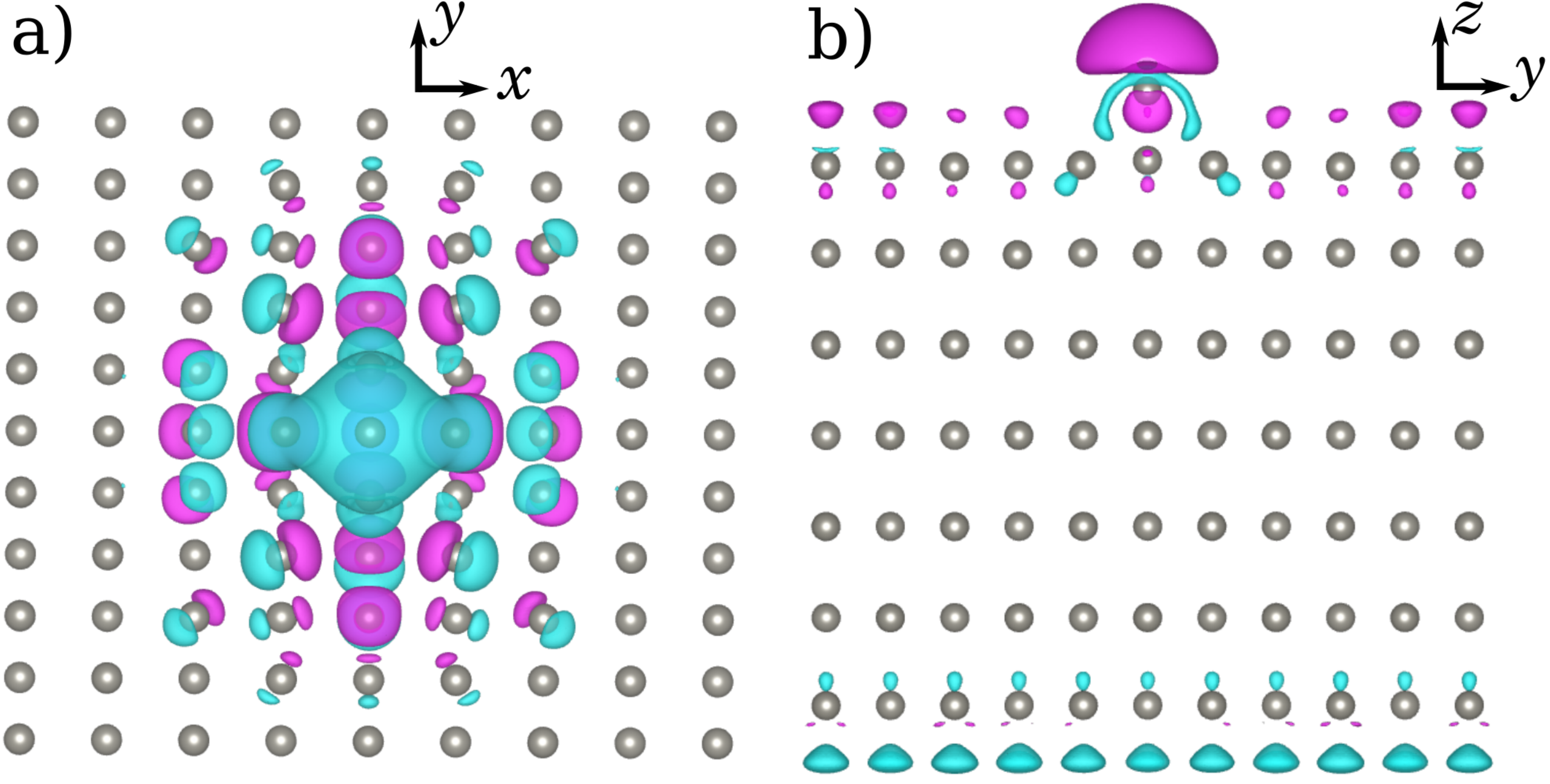}
    \caption{Charge redistribution induced by (a) the presence of an adatom, (b) a positive 1 GV/m applied field (anode) on a system with adatom (atoms fixed at their original zero-field positions for illustration purposes).
    The open surface of the slab is \{110\} oriented.
    Cyan and magenta colored areas correspond to increased and decreased electron densities, respectively, that exceed 1\% of the maximum electron density of the reference system for (a) and 0.1\% for (b).}
  \label{fig:charge_dist}
\end{figure}

An applied electric field causes charge redistribution on metal surfaces by shifting the electron densities with respect to the position of positive surface nuclei either out from the surface or into the bulk, depending on the direction of the field.
Similarly, the presence or the absence of an atom, causes charge redistribution in its vicinity as well.
Fig. \ref{fig:charge_dist} gives an illustration of how the charge distribution is modified in both cases.
This illustration shows the actual electron density obtained by our DFT calculations for a W adatom on a perfect W\{110\} surface.
It is clear that the effect of the presence of an adatom spreads well beyond its position; thus, unlike previous approaches \cite{tsong1975direct}, the change of the charge density $\rho(\vec{r})$ in the entire system has to be considered. 

The interaction of a charge distributed as $\rho(\vec{r})$ with a uniform applied electric field $\vec{F} = \hat{z}F$ changes the value of the total energy of the system.
This change can be calculated by analyzing the behavior of the corresponding system dipole moment $\vec{\mathcal{P}}=(\mathcal{P}_x,\mathcal{P}_y,\mathcal{P}_z) = \int \rho\vec{r} dV$.
An infinitesimal increment of the uniform field $\delta \vec{F}$, applied to a system with a dipole moment $\vec{\mathcal{P}}$ induces an infinitesimal change of the total energy (see sec. (4.2) in ref. \cite{jackson1975classical} or  eq. (11.3) in \cite{landau1984electrodynamics})
$\delta E = - \vec{\mathcal{P}} \cdot \delta \vec{F}$.
Since the induced system dipole $\vec{\mathcal{P}}(\vec{F})$ varies with $\vec{F}$, the energy of the system under the field is
\begin{equation} \label{eq:EofP}
	E(F) = E(0) - \int_0^F \mathcal{P}_z(F') dF' 
\end{equation}
where $E(0)$ is the total energy of the system in the absence of an external field.
For small fields, the relation $\mathcal{P}_z(F)$ can be represented as a Taylor expansion
\begin{equation} \label{eq:pofF}
	\mathcal{P}_z(F) = \mathcal{M} + \mathcal{A} F  + O(F^2) \textrm{,}
\end{equation}
where $\mathcal{M}$ is the permanent dipole moment and $\mathcal{A}$ the polarizability of the system.
Using eq. \eqref{eq:pofF} in eq. \eqref{eq:EofP} we obtain the total system energy
\begin{equation} \label{eq:Free}
	E(F) = E(0) - \mathcal{M} F - \frac{1}{2} \mathcal{A} F^2 + O(F^3) \textrm{.}
\end{equation}
See sec. (\ref{sec:meaning}) for a detailed analysis of the approximation \eqref{eq:pofF} and the physical meaning of $\mathcal{M}$ and $\mathcal{A}$ for the slab system discussed below.

Equation \eqref{eq:Free} gives the basic relation between the applied field and the total energy of the system, determined by only two parameters: the  permanent system dipole moment and system polarizability.
The latter are specific for a given system configuration, but can be found, for instance, using DFT calculations.

%

\subsection{Atomic transitions under uniform electric field} \label{sec:uniform}

Focusing on the migration and evaporation of a surface atom, we consider a rectangular metal slab under a uniform applied field, with its surfaces normal to the $z$ axis, such as the one illustrated in fig. \ref{fig:charge_dist}.
The migration energy barrier of any atom of the system is the minimum work required for its transition to a new site.
It is therefore defined as the difference in the total energy between two system configurations: one with the adatom at the saddle point (subscript $s$), where $E$ assumes the highest value along the migration path, and the other at the initial lattice site (subscript $l$).
Thus,
\begin{equation} \label{eq:Barrier_general}
E_m \equiv E_{s} - E_{l} = E_m(0) - \mathcal{M}_{\text{sl}} F - \frac{1}{2} \mathcal{A}_{\text{sl}} F^2
\end{equation}
where $E_m(0)$ is the barrier without field, $\mathcal{M}_{\text{sl}} \equiv \mathcal{M}_{s} - \mathcal{M}_{l}$ and $\mathcal{A}_{\text{sl}} \equiv \mathcal{A}_s-\mathcal{A}_l$.
In previous works \cite{giesen2005thermodynamics,muller2006migration} the energy barrier was estimated based on the two first terms of eq. \eqref{eq:Barrier_general}.
However, neglecting the system polarizability terms is not adequate in the GV/m regime discussed here.

The binding energy between an atom and the surface $E_b$ can be evaluated in a similar manner as eq. \eqref{eq:Barrier_general}.
We define $E_b$ as the work needed to move the atom sufficiently far from the surface, so that the atom-surface interaction is negligible (about 2 nm \cite{sanchez2004field}); yet, close enough that it is still under the same field $F$.
This work is the difference in total energy between the initial and final states. 
Since there is no atom-surface interaction in the final state, its total energy is the sum of the energies of the two sub-systems.
Hence, $E_b = E_r + E_a - E_l$, which yields
\begin{equation} \label{eq:Binding}
E_b = E_b(0) + \mathcal{M}_{\text{lr}} F - \frac{1}{2} (\mathcal{A}_r + \mathcal{A}_{a} - \mathcal{A}_l) F^2 \textrm{,}
\end{equation}
where the subscript $r$ denotes the reference system (substrate surface in the absence of the moving atom), $a$ denotes the isolated neutral atom and we have assumed $\mathcal{M}_{a}=0$ due to the symmetry of the free atom.

Note that $E_b$ is not the activation energy for field evaporation (ion emission under extremely high fields), but thermal evaporation of a neutral atom under a non-ionizing field.
It is therefore not to be confused with the removal work $\Lambda$, i.e. the work needed to remove an atom from the surface under field to a remote field-free space. 
Although $\Lambda$ is not the activation energy for any real physical process, it is a theoretical concept of significant importance in the field evaporation theory (FET) \cite{drechsler1960point,tsong1971measurement,forbes1982fresh}. 
Within FET, $\Lambda$ has been used under various terms such as "binding energy" \cite{drechsler1960point}, "sublimation energy" \cite{tsong1971measurement}, and "bonding energy" \cite{forbes1982fresh}.
Here we call it "removal work" in order to avoid possible confusion with our binding energy $E_b$.

$\Lambda$ coincides with $E_b$ for $F=0$, but its dependence on the field $F$ is different, due to the nature of the final state (atom under field for $E_b$).
Similarly to the case of $E_b$, in the final state the two separated systems do not interact and the total energy can be written as the sum of the subsystems $E_a(0) + E_r(F)$.
Thus the removal work $\Lambda = E_a(0) + E_r(F) - E_l(F)$ can be expressed as
\begin{equation} \label{eq:removal}
	\Lambda(F) = \Lambda(0) + \mathcal{M}_{\text{lr}} F + \frac{1}{2} \mathcal{A}_{\text{lr}} F^2
\end{equation}

We note that a quadratic (on $F$) expression for $\Lambda(F) - \Lambda(0)$ was proposed in \cite{drechsler1960point, forbes1982fresh} and a linear one in \cite{tsong1971measurement} under semi-empirical considerations.
The above form contains both linear and quadratic terms with physically well-defined coefficients that can be calculated using DFT. 
Finally, if we consider the consecutive removal of a whole layer, the mean value of the differences $\langle \mathcal{M}_\text{lr}\rangle$, $\langle \mathcal{A}_{\text{lr}} \rangle$, is the difference of $\mathcal{M},\mathcal{A}$ between the initial (with $N$ full layers) and final (with $N-1$ full layers) slab configurations, divided by the total number of atoms per layer.
In both initial and final configurations $\mathcal{M}=0$ due to symmetry.
Also, in view of eq. \eqref{eq:pol_dz}, the difference in $\mathcal{A}$ is proportional to the volume of the removed layer.
Thus, it yields $\langle \Lambda(F) \rangle = \langle \Lambda(0) \rangle + \epsilon_0 F^2 \Omega / 2$, where $\Omega$ is the atomic volume.
This result has been proven previously by Forbes \cite{forbes2001fundamental} using a fundamentally different argument.
Our agreement with Forbes offers an additional validation for our approach.

\subsection{Atomic transitions under non-uniform electric field} \label{sec:gradient}

Let us now consider the migration barrier in the presence of a small electric field gradient $\gamma \equiv dF/dx$ along the migration direction $x$.  
Such a gradient may appear due to surface features that locally enhance the applied field.
In this case, our fundamental equation \eqref{eq:Free} and the derived expression \eqref{eq:Barrier_general} is not valid directly and the estimation of migration barriers becomes more complicated. 
Nevertheless, when $\gamma$ is sufficiently small, $E_m$ can be asymptotically approximated by a formula similar to \eqref{eq:Barrier_general}.

To this end, we shall write the total energy of the system $E(\vec{r}_i)$, when the migrating atom lies at a surface point $\vec{r}_i = (x_i,y_i,z_i)$, as
\begin{equation} \label{eq:bind_general}
	E(\vec{r}_i) = E_r + \Delta E(\vec{r}_i)
\end{equation}
where  $\Delta E(\vec{r}_i)$ is the energy added to the system due to the introduction of the atom under study at $\vec{r}_i$.
$E_r$, i.e. the reference system energy in the absence of the atom, is independent of the position $\vec{r}$, therefore does not enter the expressions for the migration barrier, i.e.
\begin{equation} \label{eq:Barrier_DE}
	E_m = \Delta E(\vec{r}_s) - \Delta E(\vec{r}_l) \textrm{.}
\end{equation}
$\Delta E(\vec{r}_i)$ depends on interatomic interactions and interactions of charges with the external electric field that are localized around $\vec{r}_i$. 
In other words, only the values of $F(x)$ in the vicinity of $x_i$ affect it.
As one can see in fig. \ref{fig:charge_dist}a, the charge redistribution due to the introduction of the atom is significant only within a certain cut-off radius $R_c$, which in the simulated system does not exceed 1--2 lattice constants.

We now demand that the gradient of the field is sufficiently small so that  $\gamma R_c \ll F(x_i)$, i.e. the change of the field within $R_c$ is negligible.
If so, we can neglect any changes of the field in the area surrounding the moving atom and assume that $F(x) \approx F(x_i)$, where $F(x_i)$ is the field at the exact position $\vec{r}_i$. 
In this case, $\Delta E(\vec{r}_i)$ can be approximated by its value under the corresponding uniform field $F(x_i)$. 
If we substitute the specific points $\vec{r}_{s}$ and $\vec{r}_{l}$ in eq. \eqref{eq:Barrier_DE} and use \eqref{eq:bind_general} we obtain
\begin{equation} \label{eq:Barrier_approx}
	E_m \approx \left(E_s \left( F_s \right) - E_r \left( F_s \right) \right) - \left( E_l \left( F_l \right) - E_r \left( F_l \right) \right)
\end{equation}
with $E_i(F_i)$ being the total energy of the system in the configuration $i$ under a uniform applied field $F_i=F(x_i)$. Here $i$ stands for $s,l$ or $r$.

By combining eq. \eqref{eq:Barrier_approx} with \eqref{eq:Free}, we obtain our final formula for the migration barrier under a non-uniform electric field
\begin{equation} \label{eq:BarapproxdE}
	E_m \approx E_m(0) - \mathcal{M}_{\text{sl}} F_l - \frac{\mathcal{A}_{\text{sl}}}{2}  F_l^2 - \mathcal{M}_{\text{sr}} \Delta F - \mathcal{A}_{\text{sr}} F_l \Delta F 
\end{equation}  
where $\Delta F \equiv F(x_s) - F(x_l) = \gamma (x_s - x_l)$.
The first three terms of expression \eqref{eq:BarapproxdE} are identical to eq. \eqref{eq:Barrier_general} and give the modification of the barrier in the presence of the electric field.
The last two terms introduce the directional modifications due to the field gradient.
Similar functional forms were used by TK \cite{tsong1975direct}.
However, the physical quantities describing the field effects are different.
TK's equation is based on the atomic polarization characteristics, while our equation is derived considering the energy changes in the entire system and its polarization properties, given by the $\mathcal{M}, \mathcal{A}$ parameters.
A more detailed comparison to TK's approach will be given in a forthcoming publication.

\subsection{The slab system and the physical meaning of its polarization characteristics} \label{sec:meaning}

We shall now return to the slab model to discuss the physical meaning of its polarization characteristics and the approximations underlying its adoption.
The rectangular metal slab system introduced in section \ref{sec:uniform} is the standard system used for DFT calculations on metal surfaces \cite{neugebauer1992adsorbate,feibelman2001surface,lozovoi2003reconstruction, sanchez2004field, rusu2006work}.
The underlying approximation in such calculations (also in most atomistic simulations such as molecular dynamics and kinetic monte carlo) is that the movement of an atom is affected only by its local environment; thus a good representation of the latter is sufficient for the calculation.
The slab model provides a good representation of the local environment of an atom residing on the surface of a metal electrode of a macroscopic anode-cathode.
However, the conditions under which this approximation is valid are not usually discussed.

First, the slab has to be sufficiently thick to obtain bulk properties in its middle (i.e. $> 10-20 \text{ \AA}$); this ensures a sufficiently good representation of the effectively infinitely deep bulk below the surface.
Second, the roughness of the surface should be much smaller than the lateral width of the slab and the thickness of the vacuum region.
Third, the radius of curvature of the real surface must be much larger than the slab thickness (i.e. $> ~ 100 \text{ \AA}$).
The last two conditions ensure that the surface can be considered quasi-flat.

When the metal slab is introduced to the influence of a constant external electric field $\vec{F} = F \hat{z}$, the free charge of the metal shall redistribute in order to nullify the electric field in its interior.
Two opposite charge layers will be induced on the surfaces (see e.g. fig. \ref{fig:charge_dist}b).
We note that the applied field $F$ of the slab model corresponds to the local field of the macroscopic surface, a few nm above the considered surface atom, i.e. in a distance where the atomic movements cannot affect it, but close enough that it assumes its local value as dictated by the geometry of the macroscopic system.

It can be shown (see appendix \ref{sec:proof_A}), that the polarizability of this system can be approximated as
\begin{equation} \label{eq:pol_dz}
    \mathcal{A} \approx \epsilon_0  S \Delta z \textrm{,}
\end{equation}
where $S$ is the lateral area of the slab, $\epsilon_0$ the dielectric constant and $\Delta z$ the distance between the centers of mass of the two charge layers.
In other words, the system polarizability is proportional to the effective field-free volume inside the slab.

The above expression can be used to obtain the continuum-limit Maxwell stress.
By substituting to eq. \eqref{eq:Free} and differentiating the electrostatic energy with respect to $\Delta z$, we obtain the standard expression $\epsilon_0 F^2 / 2$ \cite{landau1984electrodynamics} for the Maxwell tensile stress on a metal surface.
We note that this represents the mean pressure exerted by the field on the slab surface. 
The total force exerted in a specific atom can be calculated similarly, but the derivatives of $\mathcal{M}, \mathcal{A}$ with respect to the atom's coordinates have to be computed separately.

From eq. \eqref{eq:pol_dz} it is evident that $\mathcal{A}$ scales linearly with the system volume and therefore cannot be considered a property of the surface or an atom. 
However, the differences in $\mathcal{A}$ such as $\mathcal{A}_{\text{sl}} \equiv \mathcal{A}_s - \mathcal{A}_l$ or $\mathcal{A}_{\text{sr}} \equiv \mathcal{A}_s - \mathcal{A}_r$, upon which characteristic transition energies depend, converge with the system size, i.e. are size-independent if the system is sufficiently large. 
This is because the differences in the charge distribution $\rho(\vec{r})$ due to a displacement of an atom are localized in its vicinity and therefore any point far from the atom would not contribute to the integral $\delta \vec{\mathcal{P}}=\int (\delta\rho) \vec{r} dV$.

The physical meaning of $\mathcal{A}_{\text{sl}}$, $\mathcal{A}_{\text{sr}}$ and $\mathcal{A}_{\text{lr}}$ emerges from the above analysis. 
They are proportional to the increase of the effective field-free volume of the metallic system due to a change on the surface. 
For a system with a given lateral area $S$, such as the one we simulated with DFT, differences in $\mathcal{A}$ are proportional to the corresponding shift of the charge layer position.

Finally, we note that when approximating $\mathcal{A}$ with $\epsilon_0 S \Delta z$, we assume that $\Delta z$ does not vary significantly with the applied field $F$. 
The latter is not in general true, since the system always responds to the application of a field and the corresponding Maxwell stress.
This response is always towards the minimization of the total energy, i.e. the increase of $\mathcal{A}$ and $\Delta z$.
Therefore, in a relaxed system, $\Delta z$ varies slightly with $F$, meaning that $\mathcal{A}$ is actually proportional to the zero-order term of a Taylor expansion of $\Delta z (F)$.
Higher order terms would contribute to the $O(F^2)$ terms in eq. \eqref{eq:pofF}, which are known as hyperpolarizability terms. 
As shown by the DFT calculations presented in sec. \ref{sec:results}, the $\mathcal{P}-F$ curve is perfectly linear within the simulated range of field and small numerical error margins.
However, when the fields approach the range of field evaporation, they might cause structural change of the surface in the vicinity of the atom under discussion and move the center of mass of the charge layer, thus introducing non-linearities.
Then higher order terms might need to be taken into account.

\section{Method} 
\label{sec:method}

The unknown $\mathcal{M}$ and $\mathcal{A}$ parameters in eqs. \eqref{eq:Barrier_general}, \eqref{eq:Binding}, \eqref{eq:removal} and \eqref{eq:BarapproxdE} can
be calculated for a specific system using DFT, which allows the full quantum mechanical calculation of the total ground-state energy of a system in the presence of an electric field \cite{neugebauer1992adsorbate}.
Furthermore, we can obtain the charge distribution in the entire system and calculate its total dipole moment by numerical integration.
Finally, the barriers, the binding energy and the removal work may be directly estimated by comparing the ground-state energies of different configurations.

\begin{figure}[!htbp]
  \centering
    \includegraphics[width=.99\linewidth]{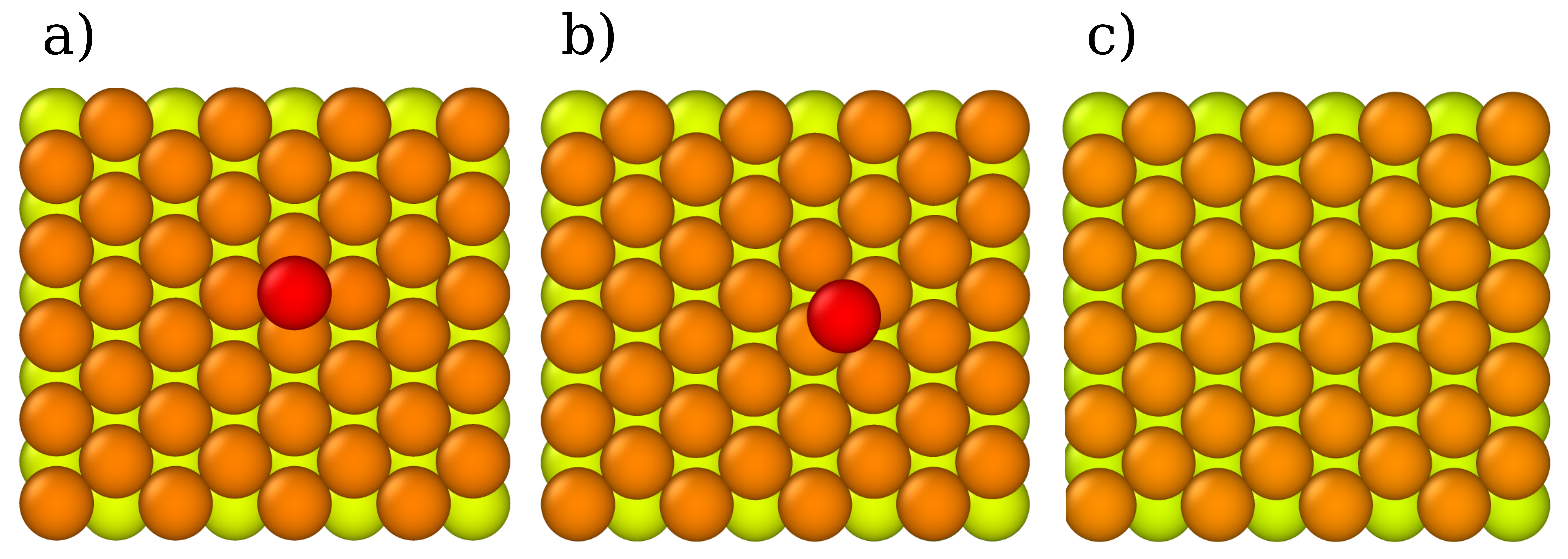}
    \caption{Top view of the slab models for the W\{110\}: (a) a slab with the W adatom at the lattice site, (b) with the W adatom at the bridge (saddle), and (c) a flat surface.}
  \label{fig:fig1_sup}
\end{figure}

Here we calculated all the parameters for the example case of a single W adatom on a flat W\{110\} surface. 
For this purpose we ran DFT simulations for four different system configurations: the flat \{110\} surface ($r$), the surface with an adatom positioned at the saddle (bridge) point ($s$), with the adatom at the lattice (hollow) point ($l$) and an isolated W atom in vacuum ($a$).
Systems ($l,s,r$) are illustrated in fig. \ref{fig:fig1_sup}(a-c) respectively.
In the cases shown in  (a) and (c), all ions were allowed to relax in all directions. 
In the saddle point case, an adatom was put in the middle of the bridge site; it was fixed in the $x$ and $y$ directions, while being allowed to relax along $z$.
We used 8 monolayers of atoms in the $x$ direction, 10 in the $y$ and 7 monolayers in the $z$ (without counting the adatom as a layer) for the adatom simulations; we shall use the notation $8 \times 10 \times 7$ for these systems. A $2 \times 2 \times 7$ system was used for the calculations of the flat surface. This minimum system with the appropriate sampling of the Brillouin zone is mathematically equivalent to any $N \times N \times 7$, for the calculation of the ground-state energy. A 24 $\textrm{\AA}$ hight vacuum was added on top of the slab for all systems. 
This vacuum height is measured from the highest fully populated atomic layer.
Finally, for the free W atom in vacuum (system $a$) a large enough box was used, so that the atom does not interact with itself over the periodic boundaries.

All our DFT calculations were performed with the Vienna ab initio simulation package (VASP) and its corresponding ultrasoft-pseudopotential database \cite{kresse1993ab,kresse1996efficiency,kresse1996efficient,vanderbilt1990soft,pasquarello1992ab,laasonen1993car,kresse1994norm}. 
VASP uses a plane wave representation for the wavefunctions.
We used the Perdew-Burke-Ernzerhof \cite{perdew1996generalized} generalized gradient approximation (GGA) functional for all calculations. 
The Blocked Davidson iteration scheme \cite{davidson1983methods} was used for the electronic relaxation and a conjugate-gradient algorithm  (see e.g. \cite{press1996numerical}) for the ionic relaxation. 
The Methfessel-Paxton smearing scheme was used to speed up the electronic relaxation \cite{methfessel1989high}. 
Finally, to avoid the charge sloshing instability, which is typical for metal slab calculations, the Kerker mixing scheme \cite{kerker1981efficient} was used. 

The electric field effect on the potential is implemented in VASP according to a scheme proposed by Neugebauer et al. in \cite{neugebauer1992adsorbate}. 
Within this scheme, an artificial dipole sheet is placed in the middle of the vacuum region that polarizes the periodic slab and introduces a uniform electric field on both sides of the slab. 
Our calculations were performed for electric fields up to 3 GV/m.
Higher fields cannot be applied with this implementation, because electrons would tunnel towards the vacuum on the cathode side of the slab, thus causing a charge sloshing that disturbs the wave functions, the total dipole moment, and the total energy of the system \cite{feibelman2001surface}. 

A Gamma-centered k-grid was used in all the calculations: a $7 \times 7 \times 1$ grid for the systems with adatom and a $28 \times 28 \times 1$ k-grid for the $2 \times 2 \times 7$ flat slab. 
The cut-off energy of the plane wave basis was set to 600 eV. 
The above values were obtained after performing convergence tests, i.e. increasing the k-grid density and the cut-off energy until the total energy of the system converged.
Our criterion for the convergence tests was 1 meV, therefore we will consider this value as our error margin for the ground-state energy calculations.
This error margin is used to obtain the error bars of all direct DFT data plotted in the figures of sec. \ref{sec:results}.

\section{Results}
\label{sec:results}

\begin{figure}[!htbp]
  \centering
    \includegraphics[width=.99\linewidth]{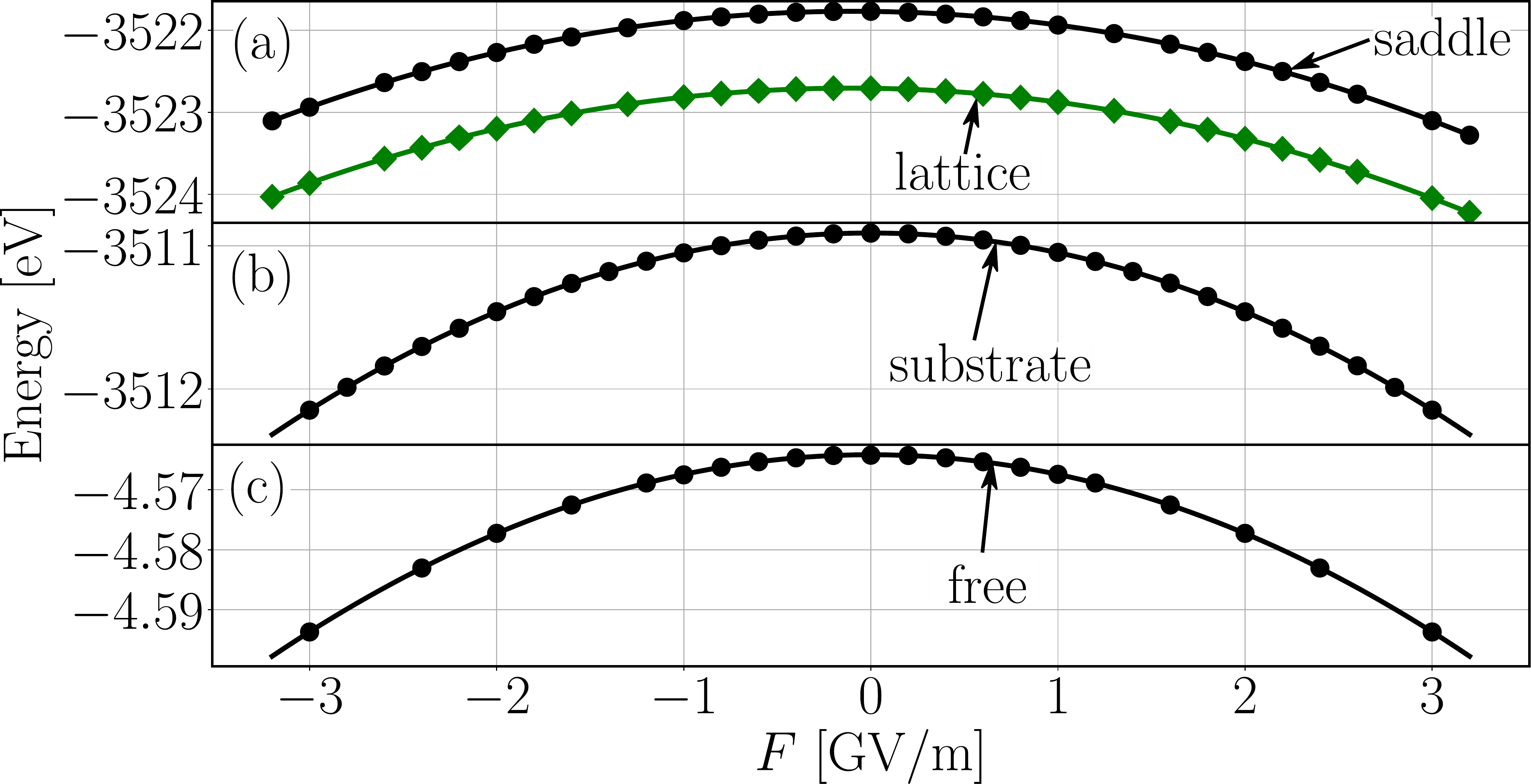}
    \caption{Total energy of the four W systems simulated by DFT, vs the applied field. Black dots and green diamonds in (a) correspond to the system with an adatom at the saddle point and at the lattice site respectively. Black dots in (b) and (c) correspond to the flat reference and isolated atom systems, respectively. The corresponding solid-line curves indicated by arrows are obtained by eq. \eqref{eq:Free} with parameters that are fitted to the same DFT data.}
    \label{fig:energy}
\end{figure}

Fig. \ref{fig:energy} shows the total energy versus the applied field as calculated by DFT (markers) for the four aforementioned systems.
We see that the DFT data follow a parabolic shape as predicted by eq. \eqref{eq:Free}.
Thus, we can obtain $\mathcal{M}$ and $\mathcal{A}$ for all systems by fitting them to the DFT data.
Table \ref{tab:tab1} summarizes the fitted parameter values along with their corresponding error estimates.
The error estimates noted as $\delta x$ for a quantity $x$ correspond to the standard deviation of the obtained value. 
The latter is obtained from the corresponding element of the least square fit covariance matrix.
As one can see in fig. \ref{fig:energy} and table \ref{tab:tab1}, the fitting results (solid lines) follow the DFT data fairly accurately with very small error margins.
\begin{table}[!htbp]
	\centering
  	\caption{System permanent dipole moment and polarizability along with their error estimations (denoted with $\delta$) as obtained by fittings to DFT data for the four simulated systems.}
  \label{tab:tab1}  
  \begin{tabular*}{\linewidth}{@{\extracolsep{\fill}} l c c c r}
	\hline \hline
    quantity 								& lattice ($l$) 			& saddle ($s$)			& substrate	($r$)		& free atom 	($a$)		\\
 	\hline
    $E(0)$ [eV]      						& -3522.707  		&  -3521.770			&  -3510.910			& -4.564048 			\\
    $\delta E(0)$ [eV]  						&$4\times 10^{-5}$ 	&  4 $\times10^{-5}$	& $5\times 10^{-4}$ 	& $5 \times 10^{-5}$	\\   
    $\mathcal{M}$  [e\AA]  							& 0.3055 			&  0.2735			& 0					&    0				\\
    $\delta \mathcal{M}$ [e\AA]     					& $10^{-4}$ 			&  $10^{-4}$   		& -					&    -				\\
	$\mathcal{A} \text{ [e\AA}^2/\text{V}]$		& 27.740  			&  27.771 			& 27.51				&	0.659			\\       
	$\delta \mathcal{A} \text{ [e\AA}^2/\text{V}]$	& $10^{-3}$			&  $10^{-3}$   		& $10^{-2}$			&   $10^{-3}$			\\
    \hline
  \end{tabular*}
\end{table} 

$\mathcal{M}, \mathcal{A}$ were extracted from the $E-F$ curves of fig. \ref{fig:energy}, although this could also be done directly from the ($\mathcal{P}-F$) curves.
In order to confirm that the fitted values of $\mathcal{M}$ and $\mathcal{A}$ correspond to the actual system permanent dipole moment and polarizability for all configurations, we compared the system dipole moment $\mathcal{P}$ (for this system due to symmetry $\vec{\mathcal{P}}=\hat{z}\mathcal{P}_z$) as calculated by numerical integration of the charge density obtained with DFT and as predicted by the linear dependence \eqref{eq:pofF} with the fitted $\mathcal{M}$ and $\mathcal{A}$ values.
We obtained a perfect agreement, with an RMS error not exceeding 0.26\% for any of the four systems.

This small deviation is attributed to the numerical error in the evaluation of $\mathcal{P}$ via integration of the electron density.
This lack of numerical precision is the very reason we chose the $E-F$ curves instead of the $\mathcal{P}-F$ ones. 
The calculation of $\mathcal{P}$ from the DFT electron density has much larger numerical error than the corresponding calculations of energy values for a given computational effort \cite{kresse2002vienna}. 
This is due to the usage of a limited number of real-space mesh grid points in our calculations and the sensitive nature of the dipole moment integral (1st order moment of a spatially oscillating quantity).
Furthermore, this effect produced an increased numerical error in the calculation of $\mathcal{P}$ for the $2 \times 2 \times 7$ flat slab system due to its decreased number of real-space mesh points.
Thus the corresponding $\mathcal{P}-F$ curve gave a significant 5\% deviation from the linear relation $\mathcal{P} = \mathcal{A}_r F$.
For this reason the dipole moment for the flat system was recalculated using a bigger system ($6 \times 8 \times 7$) for 5 field points in order to perform the comparison with the linear curve. 
The recalculated values of $\mathcal{P}$ did give a very good agreement with the linear curve, with the corresponding rms error being 0.1\%.

\begin{figure}[!htbp]
  \centering
    \includegraphics[width=.99\linewidth]{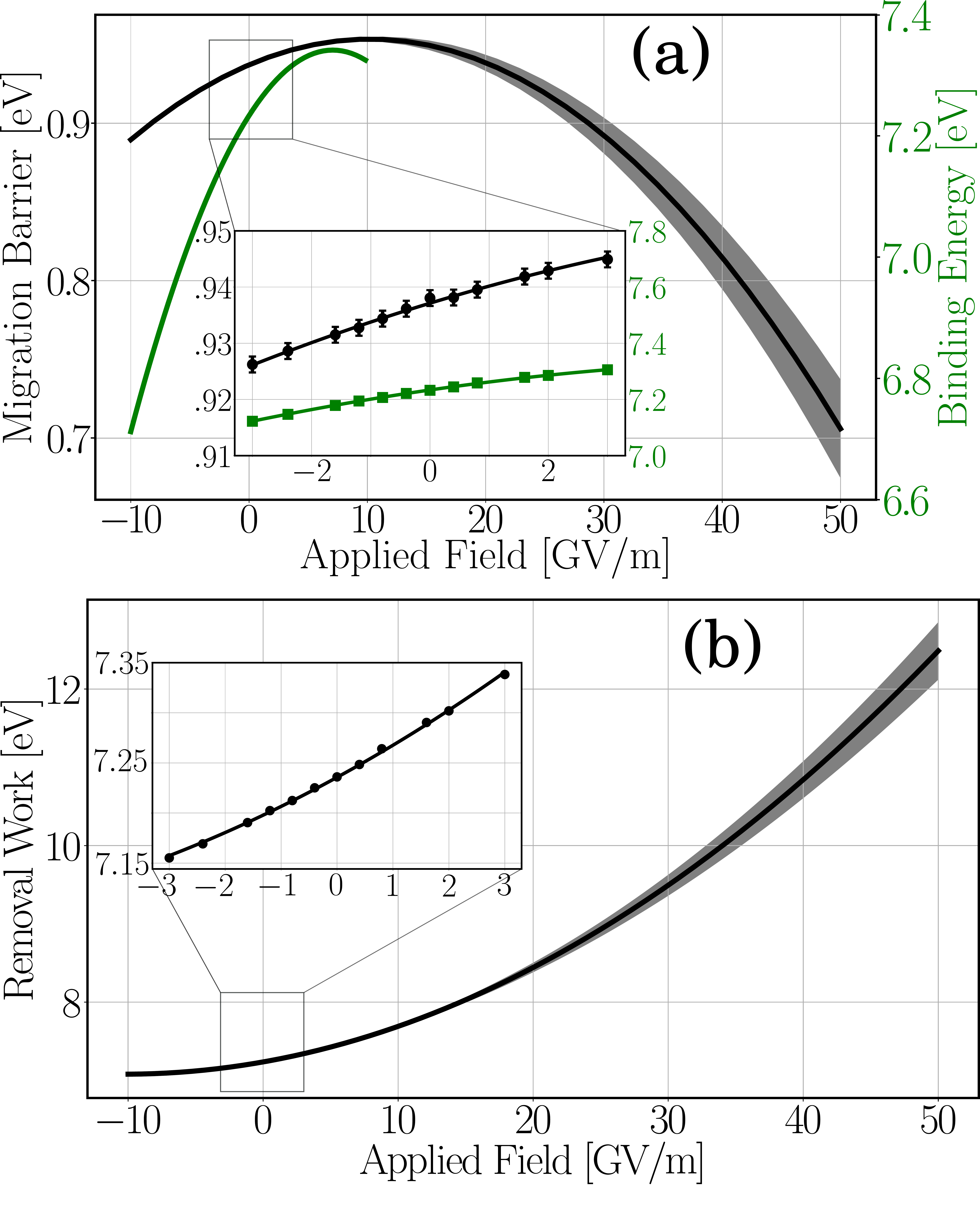}
    \caption{Migration barrier $E_m$ [sub-figure (a), black, left axis], binding energy $E_b$ [(a), green, right axis] and removal work $\Lambda$ [(b), green, right axis] of a W adatom on a W\{110\} surface vs the uniform applied electric field as calculated by eq. \eqref{eq:Barrier_approx} and \eqref{eq:Binding} respectively. 
The gray shadow area around the curves gives the error margin calculated by the error propagation rule (EPR) applied on the uncertainties of the parameters given in table \ref{tab:tab2}.
     In the insets we zoom in the low field region and plot with markers (dots for $E_m$, $\Lambda$ and squares for $E_b$) the corresponding direct DFT data. The marker errorbars are calculated by applying the EPR to the 1 meV error estimation for the DFT ground-state energy (see sec. \ref{sec:method}).} 
    \label{fig:Em_Eb_vs_F}
\end{figure}

In fig. \ref{fig:Em_Eb_vs_F}, we plot the migration barrier $E_m$, the binding energy $E_b$, and the removal work $\Lambda$ versus a uniform applied field $F$, as calculated according to eqs. \eqref{eq:Barrier_general}, \eqref{eq:Binding} and \eqref{eq:removal} respectively.
Table \ref{tab:tab2} summarizes the corresponding values of the parameters that determine $E_m, E_b \textrm{ and } \Lambda$, calculated according to the values of table \ref{tab:tab1}.
The error estimates are calculated according to the error propagation rule applied to the error margins given in table \ref{tab:tab1}.
We remind the reader that the notation $x_{\text{ab}} \equiv x_a - x_b$ denotes the difference in the value of $x$ between the system $a$ and the system $b$.
\begin{table}[!htbp]
	\centering
  	\caption{Parameters determining the migration barrier and the binding energy as a function of the applied field and the field gradient.}
  \label{tab:tab2}  
  \begin{tabular*}{.8\linewidth}{@{\extracolsep{\fill}} l c r}
	\hline \hline
    quantity 								& value 				& error 			\\
 	\hline
    $E_m(0)$ [eV]      						& 0.9371  			&  $5\times10^{-5}$		\\
    $\mathcal{M}_{\text{sl}}$  [e\AA]  						& -0.0319 			&  $2\times10^{-4}$				\\
    $\mathcal{A}_{\text{sl}} \text{ [e\AA}^2/\text{V}]$  	& 0.031 				&  $2.5\times10^{-3}$	   			\\
    $\mathcal{M}_{\text{sr}}$ [e\AA]						& 0.2735 			&  $1.3\times10^{-4}$	\\
    $\mathcal{A}_{\text{sr}} \text{ [e\AA}^2/\text{V}]$  	& 0.26 				&  $1.3\times10^{-2}$	\\
    $E_b(0) = \Lambda(0)$ [eV]      						& 7.233  			&  $5\times10^{-4}$		\\
	$\mathcal{A}_r + \mathcal{A}_a - \mathcal{A}_l \text{ [e\AA}^2/\text{V}]$		& 0.4318  			& $1.3\times10^{-2}$ 			\\
	$\mathcal{M}_\text{lr}$ [e\AA]			& 0.306  			& $1.4\times10^{-4}$ 			\\
	$\mathcal{A}_{\text{lr}} \text{ [e\AA}^2/\text{V}]$		& 0.225  			& $1.3\times10^{-2}$ 			\\      
    \hline
  \end{tabular*}
\end{table} 

In order to validate these formulae, we also calculated these values directly from $E_s$, $E_l$, $E_r$, and $E_a$ as obtained by DFT.
In the inset of fig. \ref{fig:Em_Eb_vs_F}, we compare $E_m$, $E_b$ and $\Lambda$ as obtained by the formulae (solid lines) and by DFT (markers) for the range of fields where the used DFT method can produce meaningful results.
The theoretical curves agree very well with the DFT data.

On the anode side ($F>0$), all three quantities increase for small fields, because both linear terms $-\mathcal{M}_{\text{sl}}$ and $\mathcal{M}_{\text{lr}}$ are positive.
$E_m$ and $E_b$ reach a maximum around 10 GV/m where the negative quadratic terms start dominating and the inverse trend appears.
On the other hand, both quantities are monotonously decreasing on the cathode side ($F<0$).
Therefore, any applied field would speed up the diffusion and promote evaporation on the cathode.
On the contrary, both diffusion and evaporation would slow down for an anode field up to a certain turning point.
The behavior of $\Lambda$ coincides with $E_b$ in the low field regime due to their common linear term.
Nevertheless, for higher fields $\Lambda$ exhibits an upwards curvature, due to the positive quadratic term which includes the work required to remove the atom from the influence of the field.

\begin{figure}[!htbp]
  \centering
    \includegraphics[width=.99\linewidth]{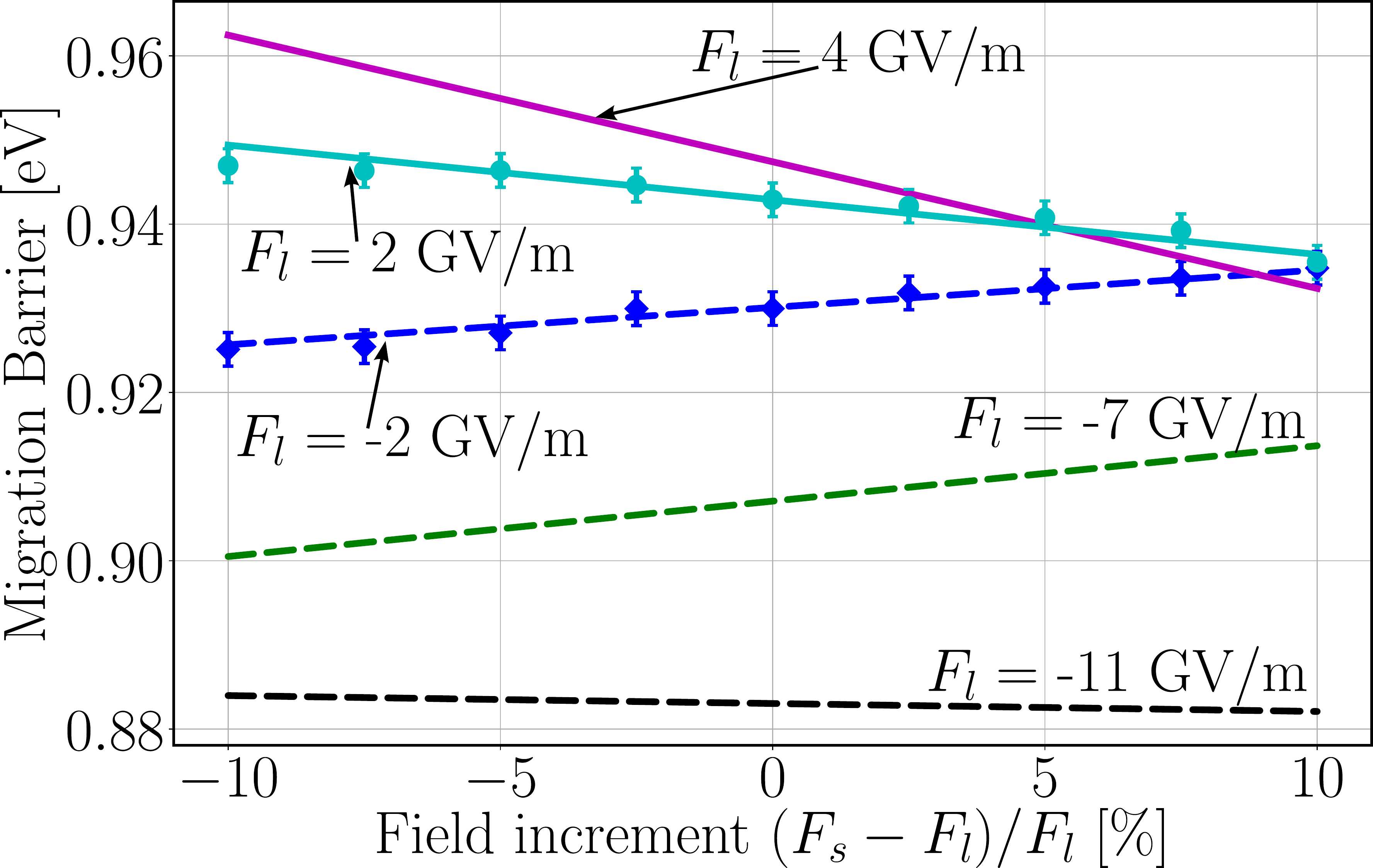}
    \caption{Migration barrier vs the relative field increment $(F_s - F_l) / F_l$ for various applied fields $F_l$.
    Solid lines correspond to the anode ($F_l>0$) and dashed ones to the cathode ($F_l<0$), as calculated by eq. \eqref{eq:BarapproxdE}.
	Markers correspond to values directly calculated by DFT according to eq. \eqref{eq:Barrier_approx}.
	The errorbars are obtained as in fig. \ref{fig:Em_Eb_vs_F}.}
    \label{fig:BarGrad}
\end{figure}

Considering the diffusion under a field gradient, which is described by equation \eqref{eq:BarapproxdE}, fig. \ref{fig:BarGrad} demonstrates both the modifications of the barrier due to the applied field and the preferable direction of the biased diffusion due to the field gradient.
Since we are here describing field differences, we plot the barrier versus the relative field increment $(F_s - F_l) / F_l$. 
Note that positive $(F_s - F_l) / F_l$ correspond to stronger fields for both the anode and the cathode.
We see that the theoretical curves are in good agreement with the direct DFT values.

In fig \ref{fig:BarGrad}, similarly as in fig \ref{fig:Em_Eb_vs_F}, a different trend appears for the anode and cathode cases, due to the linear $\mathcal{M}_{\text{sr}} \Delta F$ term in eq. \eqref{eq:BarapproxdE}.
On an anode, the diffusion is biased towards higher fields [$(F_s - F_l) / F_l > 0$], as has already been experimentally observed \cite{tsong1975direct}. 
However, counter-intuitively, for cathode fields weaker than 11 GV/m, the diffusion is preferable towards weaker fields [$(F_s - F_l) / F_l < 0$].
Nevertheless, the bias (indicated by the corresponding line slope) is much weaker than for the anode case and it weakens further as the applied cathode field increases. 
Above 11 GV/m fields, the cathode migration energy follows the same trend as for the anode, i.e. the diffusion is again biased towards stronger fields.
This turning point depends on the equilibrium between the fourth and fifth term of equation \eqref{eq:BarapproxdE}.

The trends shown in fig. \ref{fig:Em_Eb_vs_F} and \ref{fig:BarGrad} are determined by the balance between 1st and 2nd order terms in eqs. \eqref{eq:Barrier_general}, \eqref{eq:Binding}, \eqref{eq:removal} and \eqref{eq:BarapproxdE}.
From a physical point of view, the 1st order terms correspond to $\mathcal{M}$-values, i.e. the permanent dipole moment due to the adatom-induced charge redistribution, such as the one shown in fig. \ref{fig:charge_dist}(a).
On the other hand, 2nd order terms depend on how the field-induced charge redistribution [see fig. \ref{fig:charge_dist}(b)] is modified in different configurations.

Note that the $\mathcal{M}$ and $\mathcal{A}$ coefficients of these terms might differ significantly for different migration processes and materials.
Hence, our results for the simple W-on-W\{110\} are not enough to draw general conclusions on the diffusion on more complex surfaces.
In order to do that, $\mathcal{M}$ and $\mathcal{A}$ should be calculated for additional atomic migration processes.
We leave this out of the scope of this work, which focuses on the theoretical framework upon which such calculations may be based.

\section{Discussion} \label{sec:discussion}

\subsection{Experimental validation} \label{sec:experiment}

We chose to perform DFT calculations for the W\{110\} system because experimental data of the diffusion on this specific system are available for comparison.
Tsong and Kellogg \cite{tsong1975direct} conducted experiments of the biased diffusion of adatoms under a non-uniform field. 
They cut a W tip on its \{110\} surface and then placed an adatom on it.
The position of the latter was monitored by field ion microscopy.
Thus, they observed and measured the brownian motion of the adatom on the surface both in the presence and the absence of an applied field.

If the barrier is considered to depend linearly on the field gradient as in eq. \eqref{eq:BarapproxdE}, a straight-forward dependence of the coefficient of $\Delta F$, $B \equiv \mathcal{M}_{\text{sr}} + \mathcal{A}_{\text{sr}}F$, to directly measurable quantities can be derived (see appendix \ref{sec:exp}).
It yields
\begin{equation} 
	\label{eq:bias}
	B  = \frac{2kT}{l\gamma} \sinh^{-1} \left( \frac{l \langle x \rangle_b }{2\langle x^2 \rangle} \right) \textrm{,}
\end{equation}
where $\langle x \rangle_b$ is the mean displacement of the adatom when it performs a biased diffusion under a non-uniform field with gradient $\gamma$, $\langle x^2 \rangle$ is the mean square displacement of the adatom when the latter diffuses without any applied field, $l$ is the atomic jump length, $k$ is the Boltzmann constant and $T$ the temperature.

Now in ref. \cite{tsong1975direct} both $\langle x \rangle_b$ and $\langle x^2 \rangle$ were measured on the same surface under the same conditions.
$\langle x^2 \rangle$ was measured without a field, whereas a field was applied to measure the biased displacement $\langle x \rangle_b$.
By inserting the measurements in eq. \eqref{eq:bias}, Tsong and Kellogg obtained $B = 1.14 \textrm{ e\AA}$.
Their applied field value was estimated $F= \textrm{23.5 GV/m}$ and the corresponding field gradient $\gamma = 0.0134 \textrm{ V/\AA}^2$ (this value is multiplied by the correction factor $\sqrt{3}/2$ to account for the misalignment of the jump direction on the hexagonal lattice with the field gradient).
From our extracted values we obtain $\mathcal{M}_{\text{sr}} + \mathcal{A}_{\text{sr}} F = 0.88 \pm 0.03 \textrm{ e\AA}$.
This result is in a surprisingly good agreement with the experimental value, with a small deviation of about 20\%.

This small deviation can be attributed to the experimental error of the involved measurements.
In table I of ref. \cite{tsong1975direct}, the reported error of the extracted $B$ value ($\alpha$ in \cite{tsong1975direct}) is 11.4\%.
However, a closer examination of TK's results shows that the experimental error margin should be significantly higher.
Although no details are given by TK on their method of obtaining the reported error margin in $B$, a minimum error can be estimated already by their figure 5.
In the latter they report direct measurements of $\langle x^2 \rangle$, with errorbars that are not less than 25\% for each measurement.
By neglecting any other possible source of error ($\langle x \rangle_b, \gamma \text{ and } kT$ most probably also have significant error margins) and applying the error propagation rule to \eqref{eq:bias}, we obtain a 21\% error margin in the experimental value of $B$, meaning that there is a quite reasonable agreement between experiment and theory.

\subsection{Limits of the theory}

The range of fields used for the theoretical curves in fig. \ref{fig:Em_Eb_vs_F} is not arbitrary.
It is rather dictated by the limits of validity of the assumptions used to develop the current model.
For $F<-10$ GV/m (cathode fields higher than 10 GV/m), field emission becomes significant and the space charge may affect the dynamics of the field distribution around the defect.
Furthermore, due to the intense field emission initiating vacuum arcs, such fields are almost impossible to realize experimentally without causing an instant vacuum breakdown \cite{Dyke1953I} and are therefore of limited interest.

On the other hand, high positive fields also impose strict limitations due to the fundamental definitions of the migration barrier and the binding energy we use here.
The definition of the binding energy as $E_b = E_r + E_a - E_l$ assumes that if one moves the adatom away from the surface under field, the total energy of the system will converge to $E_r + E_a$ as the distance increases. 
However, at the fields approaching the field evaporation regime (30-60 GV/m for W) \cite{miller2014atom}, the above assumption is not valid.
In this case the applied field distorts significantly the potential "well" of the atoms, thus causing tunnelling of electrons from the atom towards the metal slab, even for high distances between the former and the latter.
This means that the atom and the slab get partially charged, which adds a significant distance-dependent component on the total energy.
Therefore the system energy decreases linearly with the atom-slab distance instead of converging, as we have assumed in our model. 
A detailed DFT calculation revealing this behavior for Al surfaces can be found in ref. \cite{sanchez2004field}.

To investigate the field range where this behavior is expected to appear for W, we ran DFT simulations for an isolated W atom under high fields.
Already at $F=10$ GV/m the system energy deviates significantly from the parabolic behavior shown in fig. \ref{fig:energy}c and the wave functions are non-zero in the vacuum region of the system.
For this reason we limit the plot of the binding energy at below 10 GV/m.

The migration barrier, on the other hand, becomes rather meaningless when the activation energy for the field evaporation becomes comparably small or even smaller.
This is because the atoms will see a potential "slide" towards the vacuum before they reach the new site.
The evaporation activation energy has been measured to be 0.9 eV for a field of 47 GV/m \cite{kellogg1984measurement}. 
According to our calculations, the migration barrier at this field is about 0.75 eV, which is close to the evaporation activation energy.
Therefore, we limit our calculation for the barrier at fields up to 50 GV/m.

Finally, the limited range of the electric fields ($|F|$ less than 3--4 GV/m in our calculations) that can be calculated by the current DFT technique \cite{feibelman2001surface} also affects the precision of the model at higher fields.
Although the qualitative description of our theory can be considered valid up to 40--50 GV/m, the quantitative results might become inaccurate already at lower fields. 
This is because the error margins in the calculation of the polarization parameters $\mathcal{M}$ and $\mathcal{A}$ are enough to give an significantly increasing uncertainty at high fields, as is evident from the increasing error bars of fig. \ref{fig:Em_Eb_vs_F}.
Furthermore, an additional uncertainty originates from the fact that second or higher order terms in eq. \eqref{eq:pofF} might become significant at high fields and therefore equations \eqref{eq:Barrier_general}, \eqref{eq:Binding}, \eqref{eq:removal} and \eqref{eq:BarapproxdE} need to be corrected with third and higher order terms as well.

\section{Conclusions}

To conclude, this work provides a rigorous theoretical basis for understanding the atomistic behavior of metal surfaces under high electric fields and can be used to develop atomistic computational models for the long-term evolution of metal surfaces in this condition.
We have showed that the behavior of a surface atom can be described with a few parameters in terms of the total dipole moment of both the permanent and  field-induced charges in its vicinity.
Our theory is in excellent agreement with DFT calculations and when we combine the two, we obtain results on the behavior of W adatoms on W\{110\} surfaces that are in very good agreement with experiments.

\section*{Acknowledgements}
A. Kyritsakis was supported by the CERN K-contract (No. 47207461), E.\;Baibuz by the CERN K-contract and the doctoral program MATRENA of the University of Helsinki, and V.\;Jansson by the Academy of Finland (Grant No.\;285382) and Waldemar von Frenckells Stiftelse. F.\;Djurabekova acknowledges gratefully the financial support of the Academy of Finland (Grant No. 269696). We also acknowledge the grants of computer capacity from the Finnish Grid and Cloud Infrastructure (persistent identifier urn:nbn:fi:research-infras-2016072533).
A.K. and E.B. would like to thank Dr. Ali Akbari for sharing his expertise on VASP. 

\appendix




\section{Proof of eq. \eqref{eq:pol_dz}} \label{sec:proof_A}

Let us consider a rectangular metal slab, with its top and bottom surfaces perpendicular to the $z$ direction.
If this system is introduced to the influence of a constant external electric field $\vec{F} = F \hat{z}$, the free charge of the metal will redistribute in order to nullify the electric field in its interior.
Thus two opposite charge layers at the top and the bottom of the slab will be formed. 

The total charge per area in the top layer is
\begin{equation} \label{eq:sigma_def}
	\sigma_+ = \frac{1}{S} \int_{\Omega_+} ~\rho dV
\end{equation}
where $\Omega_+$ denotes the top half volume of the slab, $\rho$ is the local charge density and $S$ is the surface area of the $x-y$ plane of the slab.
The corresponding bottom layer $\sigma_-$ can be defined equally for $\Omega_-$.
By applying the Gauss law we obtain
\begin{equation} \label{eq:sigma_i}
	\sigma_+ = - \sigma_- =  F \epsilon_0 
\end{equation}
where $\epsilon_0$ is the dielectric permittivity of vacuum.

The center of mass of the charge layers can be defined as
\begin{equation} \label{eq:rcm}
	\vec{r}_{\text{cm}}^{(i)} = \frac{\int_{\Omega_i} ~\rho \vec{r}dV}{\int_{\Omega_i} ~\rho dV}
\end{equation}
where $i$ can be either $(+)$ or $(-)$. 
If we calculate now the total dipole moment of the system, we obtain
\begin{equation} \label{eq:ptot_sm}
    \vec{\mathcal{P}} \equiv \int_{\Omega} ~\rho \vec{r} dV = \vec{\mathcal{P}}(F=0) +  S \left( \sigma_{+} \vec{r}_{\text{cm}}^{(+)} - \sigma_- \vec{r}_{\text{cm}}^{(-)} \right)  \textrm{.}
\end{equation}
The $z$ component of the dipole moment which determines the energy can then be expressed as 
\begin{equation} \label{eq:pz_sm}
    \mathcal{P}_z = \mathcal{M} + S \epsilon_0 \Delta z F \approx \mathcal{M} + \mathcal{A} F
\end{equation}
where $\Delta z = z_{\text{cm}}^{(+)} - z_{\text{cm}}^{(-)}$ is the vertical distance between the centers of mass of the charge layers.
It is evident that under the approximation that $\Delta z$ does not depend on $F$, the system polarizability is approximately  $\mathcal{A} \approx \epsilon_0 S \Delta z$.

\section{Derivation of eq. \eqref{eq:bias}} \label{sec:exp}

According to the brownian motion theory, when jumps in all directions are equally probable with an activation energy $E_m$, the mean displacement after a time $\tau$ is $\langle x \rangle = 0$ and the mean square displacement is 
\begin{equation}
	\langle x^2 \rangle = \nu \tau l^2 \exp \left(- \frac{E_m}{kT} \right)
\end{equation}
where $\nu$ is the attempt frequency, $l$ is the length of the jump and $kT$ is the temperature multiplied by the Boltzmann constant.

On the other hand, if the activation energy is $E_m+\delta E_m$ on the left side and $E_m-\delta E_m$ on the right, then the mean displacement of the biased diffusion is
\begin{equation}
	\langle x \rangle_b = 2\nu \tau l \exp \left(- \frac{E_m}{kT} \right) \sinh \left(- \frac{\delta E_m}{kT} \right) \textrm{.}
\end{equation}
The ratio between them is then
\begin{equation}
	\frac{\langle x \rangle_b }{\langle x^2 \rangle} = \frac{2}{l} \sinh \left(- \frac{\delta E_m}{kT} \right) \textrm{.}
\end{equation}
If the bias is due to an applied field that has a gradient in a certain direction, as we assumed in our theory, we can substitute $\delta E_m$ by the function of the field and the field gradient given in eq. (9), i.e.
\begin{equation}
	\delta E_m = \left( \mathcal{M}_{\text{sr}} + \mathcal{A}_{\text{sr}} F \right) \Delta F = \left( \mathcal{M}_{\text{sr}} + \mathcal{A}_{\text{sr}} F \right) \gamma \frac{l}{2} \text{.}
\end{equation}
Then the ratio becomes
\begin{equation}
	\frac{\langle x \rangle_b }{\langle x^2 \rangle} = \frac{2}{l} \sinh \left(- l\gamma \frac{\mathcal{M}_{\text{sr}} + \mathcal{A}_{\text{sr}} F}{2kT} \right) \textrm{,}
\end{equation}
which means that the "bias coefficient" $B \equiv \mathcal{M}_{\text{sr}} + \mathcal{A}_{\text{sr}}F$ can be expressed as a function of directly measurable quantities, i.e. 
\begin{equation} 
	\label{eq:biased}
	B = \frac{2kT}{l\gamma} \sinh^{-1} \left( \frac{l \langle x \rangle_b }{2\langle x^2 \rangle} \right) \textrm{,}
\end{equation}
which identical to eq. \eqref{eq:bias}.

\bibliography{bibliography/abbreviated}

\end{document}